\documentclass[letterpaper,twocolumn,10pt]{article}
\usepackage{usenix2019_v3}
\usepackage{tugraz_defaults}
\usepackage{multirow}

\newcommand{\ToolName}{{Transynther}\xspace}

\newcommand{\para}[1]{\smallskip\noindent\textbf{{#1.}}}

\makeatletter
\newcommand\footnoteref[1]{\protected@xdef\@thefnmark{\ref{#1}}\@footnotemark}
\makeatother

\usepackage{hyperref}
  \definecolor{linkcolor}{rgb}{0,0,0.25}
  \definecolor{citecolor}{rgb}{0,0.4,0}
  \definecolor{urlcolor}{rgb}{0,0,0.65}
  \hypersetup{pdfpagemode=UseNone,pdfstartview=FitH,colorlinks=true, linkcolor=linkcolor, urlcolor=urlcolor, citecolor=citecolor}

\definecolor{dkgreen}{rgb}{0,0.6,0}
\definecolor{gray}{rgb}{0.5,0.5,0.5}
\definecolor{mauve}{rgb}{0.58,0,0.82}
\author{Anonymous Submission}

\usepackage{authblk}
\author[]{Daniel Moghimi}
\affil[]{Worcester Polytechnic Institute, Worcester, MA, USA}

\newlength\figureheight
\newlength\figurewidth

\begin{document}

\title{Data Sampling on MDS-resistant 10th Generation Intel Core (Ice Lake)}

\maketitle

%==============================================================================
% Abstract
%==============================================================================
\begin{abstract}
Microarchitectural Data Sampling (MDS) is a set of hardware vulnerabilities in Intel CPUs that allows an attacker to leak bytes of data from memory loads and stores across various security boundaries. 
On affected CPUs, some of these vulnerabilities were patched via microcode updates. 
Additionally, Intel announced that the newest microarchitectures, namely Cascade Lake and Ice Lake, were not affected by MDS. 
While Cascade Lake turned out to be vulnerable to the ZombieLoad v2 MDS attack (also known as TAA), Ice Lake was not affected by this attack. 

In this technical report, we show a variant of MSBDS (CVE-2018-12126), an MDS attack, also known as Fallout, that works on Ice Lake CPUs. 
This variant was automatically synthesized using \ToolName, a tool to find new variants of Meltdown-type attacks. 
Based on the findings of \ToolName, we analyze different microcodes regarding this issue, showing that only microcode versions after January 2020 prevent exploitation of the vulnerability. 
These results show that \ToolName is a valuable tool to find new variants, and also to test for regressions possibly introduced with microcode updates. 

\end{abstract}

\section{Introduction}
In May 2019, a class of CPU attacks, under the umbrella of microarchitectural data sampling (MDS), showed that Meltdown-style attacks are still possible on Intel Core microarchitecture~\cite{Canella2019Fallout,
VanSchaik2019RIDL,Schwarz2019ZombieLoad,vanbulck2018foreshadow}.
MDS fundamentally exploits the same root cause as Meltdown~\cite{Lipp2018meltdown} and Foreshadow~\cite{vanbulck2018foreshadow}.
MDS attacks showed that Meltdown-type attacks are not limited to the L1 cache but that they can generally leak stale data from various internal CPU buffers~\cite{moghimi2020medusa}.
In contrast to Meltdown~\cite{Lipp2018meltdown} or Foreshadow~\cite{vanbulck2018foreshadow}, the attacker has only limited control of what data can be leaked, as these internal buffers only contain \emph{memory loads} and \emph{stores} from different execution contexts, regardless of any architectural security boundary. 

Fallout~\cite{Canella2019Fallout}, or microarchitectural store buffer 
data sampling (MSBDS), is an MDS variant leaking recent data stores from the store buffer.
MSBDS has been shown in attacks on the kernel space as well as on SGX~\cite{Canella2019Fallout}.
In addition to the leakage component, MSBDS can also be turned around easily to transiently inject values into a victim, an attack technique known as Load Value Injection (LVI)~\cite{van2020lvi}.

To mitigate these vulnerabilities, Intel released software workarounds and microcode updates for affected CPUs~\cite{IntelMDS}.
One of the most recent microarchitectures, Ice Lake, is reported to be unaffected by MDS attacks~\cite{ProtectionCPUModel,AffectedCPUModel}. 
Intel has also explicitly listed Ice Lake processors as not being vulnerable to LVI-SB~\cite{van2020lvi}, which exploits MSBDS for Load Value Injection~\cite{LVICPUModel}.

To better analyze the MDS vulnerabilities, and potentially find new variants, Moghimi~\etal\cite{moghimi2020medusa} presented \ToolName. 
\ToolName mutates the basic primitives of existing Meltdown-type attacks to automatically generate and evaluate new subvariants. 
In the original publication~\cite{moghimi2020medusa}, \ToolName did not only reproduce existing attacks but also discovered a 
ZombieLoad~\cite{Schwarz2019ZombieLoad} variant (Medusa) that targets the write-combining mechanisms.
In this report, we show that \ToolName also found a variant of MSBDS that works on the allegedly unaffected Ice Lake microarchitecture.

\para{Responsible Disclosure}
We have reported this finding to the Intel Product Security Incident 
Response Team (iPSIRT) on March, 27, 2020.
On May 5 2020, iPSIRT completed the triage of our proof of concept, and
they replied that the mitigation for MSBDS was not ported correctly to the 
Ice Lake microarchitecture.
As a result, Ice Lake required a microcode patch, which they developed as 
part of their late November 2019 microcode version 0x5C.
In the May 2020 update of Intel's specification update for the 10th Generation 
Intel Core Processor Family, a new errata, 057, has been added.
This errata mentions that the \texttt{MDS\_NO} bit in \texttt{IA32\_ARCH\_CAPABILITIES} control 
registers were incorrectly set~\cite{errataIceLake}. 
Intel requested an embargo until July 14, 2020, to allow enough time for OEMs 
and their customers to deploy these patches.
Intel credited us by updating the advisory regarding Microarchitectural Data 
Sampling on July 14, 2020~\cite{mdsadv}. 
The report was awarded a bug bounty.

\section{Fallout (MSBDS) on Ice Lake}
In this section, we present the MSBDS variant automatically synthesized by \ToolName~\cite{moghimi2020medusa}.
This MSBDS variant is the only known MDS attack that works on the Ice Lake microarchitecture. 
We analyze the leakage rates of this variant, and also the affected microcode versions. 

\para{Discovery}
We ran \ToolName on a Core i5-1035G1 CPU with the latest microcode that is 
shipped with Ubuntu 18.04, version 0x48. 
After running for about \SIx{5000} iterations, \ToolName reported store-to-load-forwarding leakage due to 4K aliasing of store addresses with a faulty memory load. 
This behavior was initially exploited in Fallout to bypass KASLR and leak cryptographic 
keys from the kernel space~\cite{Canella2019Fallout}. 

Based on the generated proof-of-concept, we produced a minimal working example to analyze the auto-generated proof of concept that triggers MSBDS manually. 

\para{Classification}
We noticed that MSBDS on Ice Lake only works with memory load operations
that suffer a permission failure due to accessing privileged memory (cleared US bit), 
or accessing a memory page with wrong protection keys~\cite{Canella2019}. 
Based on the systematization of Canella et al.~\cite{Canella2019}, and to the best 
of our knowledge, we conclude that Ice Lake is only vulnerable to Meltdown-US 
or Meltdown-MPK attacks. 

\para{Modified Cache State}
One of the observations from \ToolName is that the leakage rate increases significantly if the target store address is flushed from the cache.
We observe the same behavior for other instructions that modify the 
cache state. 
Specifically, executing \texttt{lock incl} on the store address leads to an even higher leakage rate than flushing the store address using \texttt{clflush}. 
\Cref{lst:gcryptecdsamasking} shows our simplified proof of concept that demonstrates MSBDS on the Ice Lake microarchitecture.
Uncommenting Line 16 or 17 modifies the cache state of the store address, resulting in a faster leakage.
If we do not modify the cache state, we observe a very slow leakage of approximately \SI{1}{\byte/\second}. 
As we can see in~\cref{tab:results}, with approximately \SI{750}{\byte/\second}, the leakage rate is significantly higher when using \texttt{lock inc} instruction (Line 19) to modify the cache state.

\begin{listing}[ht!]
\begin{lstlisting}[basicstyle=\scriptsize, numbers=left, stepnumber=1]
unsigned int c = 0;
while(cond){ 
    if(!setjmp(trycatch_buf))
        s_faulty_load();
        
    for (size_t i = 1; i < 256; i++)
        if (flush_reload((uint8_t *)&oracles + i*PAGE_SIZE)) 
            if(i == 0x41) c++;
}

/** asm.S **/
    .global s_faulty_load
s_faulty_load:
 lea address_normal+0x4822, %r14    // Store address 
 lea address_supervisor+0x822, %r15 // Load address (4K alias)  
//clflush (%r14)                    // Uncomment to modify cache state
//lock;incl (%r14)                  // Uncomment to modify cache state
 movb $0x41, (%r14)                 // Store
 movb (%r15), %al                   // Faulty Load
 lea oracles, %r13                  // Encode
 and $0xff, %rax
 shlq $12, %rax
 movb (%r13,%rax,1), %al
 ret
\end{lstlisting}
  \caption{
    Proof of concept for MSBDS on IceLake.
  }
\label{lst:gcryptecdsamasking}
\end{listing}

We also tested the proof-of-concept on various microcode versions on the Ice Lake CPU.
As not all issued microcodes are officially available by CPU vendors, we used a 
crowd-sourced repository of available microcodes~\cite{mcs}.
For our analysis, we applied 10 different compatible microcode versions, i.e., microcodes that match the CPUID of our target CPU. 
As we can see in~\cref{tab:results}, all Intel microcodes until mid-November are 
vulnerable to MSBDS, although Ice Lake should fundamentally be resistant against 
MDS attacks.

\begin{table}[htb]
\centering
\caption{List of tested microcodes on a Core i5-1035G1 CPU. For vulnerable microcodes,
the leakage rate is much higher if the target store is in a modified state, as 
it is shown by using cache flush and modification instructions. 
We ran each experiment for two minutes.}
\resizebox{1\hsize}{!}{
\begin{tabular}{lllrrr}
\toprule
\multirow{2}{*}{\rotatebox{20}{\textbf{MC Version}}} & \multirow{2}{*}{\rotatebox{20}{\textbf{MC Date}}} & \multirow{2}{*}{\rotatebox{20}{\textbf{Vulnerable}}} & \multicolumn{3}{c}{\textbf{Leakage (bytes/s)}}                            \\
                                      &                                   &                                      & \textit{\textbf{clflush}} & \textit{\textbf{lock inc}} & \textbf{Unmodified} \\\midrule
0x32 (stock)                         & 2019-07-05                        & \cmark                & 577.87                    & 754.99                     & 1.58          \\
0x36                                 & 2019-07-18                        & \cmark                & 148.24                    & 529.84                     & 0.62          \\
0x46                                 & 2019-09-05                        & \cmark                & 130.15                    & 695.80                     & 0.11          \\
0x48                                 & 2019-09-12                        & \cmark                & 271.69                    & 620.07                     & 0.59          \\
0x50                                 & 2019-10-27                        & \cmark                & 96.54                     & 542.10                     & 0.25          \\
0x56                                 & 2019-11-05                        & \cmark                & 145.46                    & 751.40                     & 0.08          \\
0x5a                                 & 2019-11-19                        & \cmark                & 532.40                    & 645.32                     & 0.70          \\
0x66                                 & 2020-01-09                        & \xmark                & 0                         & 0                          & 0             \\
0x70                                 & 2020-02-17                        & \xmark                & 0                         & 0                          & 0             \\
0x82                                 & 2020-04-22                        & \xmark                & 0                         & 0                          & 0             \\
0x86                                 & 2020-05-05                        & \xmark                & 0                         & 0                          & 0             \\\bottomrule
\end{tabular}
}
\label{tab:results}
\end{table}

\section{Conclusion}
Our report shows that Intel Ice lake client processors with early firmware version 
are vulnerable to MDS attacks. In discussions with Intel engineers we were told that 
MSBDS mitigations are present in hardware, but where disabled in early versions of 
these processors. 
It is crucial for OEMs and users to apply these latest microcode updates to 
enable protection against MDS attacks.

More importantly, our analysis using \ToolName highlights the importance 
of automated vulnerability testing and analysis for hardware and 
microarchitectural vulnerabilities.
Although \ToolName is an academic prototype, it stills proved to be a valuable tool 
for automated testing of hardware. 
The newly reported MSBDS vulnerability would not have gone unnoticed on Ice Lake,
if the earlier prototypes of the hardware were tested using such tools. 
Furthermore, OEMs could have tested these vulnerabilities before shipping 
consumer laptop with a vulnerable microcode update.

\bibliographystyle{plain}
{\footnotesize
  \bibliography{ref}
}

\end{document}